\documentclass[%
 aip,
 amsmath,amssymb,
 reprint,%
]{revtex4-1}

\usepackage{graphicx}
\usepackage{dcolumn}
\usepackage{bm}

\usepackage[utf8]{inputenc}
\usepackage[T1]{fontenc}
\usepackage{mathptmx}

\usepackage{xcolor}
\usepackage{soul,comment}

\begin{document}

\preprint{AIP/123-QED}

\title[Effect of Dimensionality on the Optical Absorption Properties of CsPbI$_3$ Perovskite Nanocrystals]{Effect of Dimensionality on the Optical Absorption Properties of CsPbI$_3$ Perovskite Nanocrystals}

\author{Albert Liu}
\affiliation{Department of Physics, University of Michigan, Ann Arbor, Michigan, USA}

\author{L. G. Bonato} 
\affiliation{Instituto de Quimica, Universidade Estadual de Campinas, Campinas, Sao Paulo, Brazil}

\author{Francesco Sessa}
\affiliation{Department of Physics, University of Michigan, Ann Arbor, Michigan, USA}

\author{Diogo B. Almeida}
\affiliation{Department of Physics, University of Michigan, Ann Arbor, Michigan, USA}
\affiliation{Instituto de Fisica, Universidade Estadual de Campinas, Campinas, Sao Paulo, Brazil}

\author{Erik Isele}
\affiliation{Department of Electrical Engineering and Computer Science, University of Michigan, Ann Arbor, Michigan, USA}

\author{G. Nagamine}

\author{L. F. Zagonel} 
\affiliation{Instituto de Fisica, Universidade Estadual de Campinas, Campinas, Sao Paulo, Brazil}

\author{A. F. Nogueira} 
\affiliation{Instituto de Quimica, Universidade Estadual de Campinas, Campinas, Sao Paulo, Brazil}

\author{L. A. Padilha}
\email{padilha@ifi.unicamp.br}
\affiliation{Instituto de Fisica, Universidade Estadual de Campinas, Campinas, Sao Paulo, Brazil}

\author{Steven T. Cundiff}
\email{cundiff@umich.edu}
\affiliation{Department of Physics, University of Michigan, Ann Arbor, Michigan, USA}

\date{\today}

\begin{abstract}
    The band-gaps of CsPbI$_3$ perovskite nanocrystals are measured by absorption spectroscopy at cryogenic temperatures. Anomalous band-gap shifts are observed in CsPbI$_3$ nanocubes and nanoplatelets, which are modeled accurately by band-gap renormalization due to lattice vibrational modes. We find that decreasing dimensionality of the CsPbI$_3$ lattice in nanoplatelets greatly reduces electron-phonon coupling, and dominant out-of-plane quantum confinement results in a homogeneously broadened absorption lineshape down to cryogenic temperatures. An absorption tail forms at low-temperatures in CsPbI$_3$ nanocubes, which we attribute to shallow defect states positioned near the valence band-edge.
\end{abstract}

\maketitle

\section{Introduction}

Colloidal nanocrystals, following decades of extensive study, have begun maturing as a material platform for commercial applications such as displays \cite{Choi2018} and photovoltaics \cite{Yuan2016}. However, despite more than 30 years of research into alternative material platforms, the initial chalcogenide-based colloidal nanocrystals have remained superior in both performance and stability for practical devices. Recently, synthesis of cesium lead-halide perovskite nanocrystals was achieved \cite{Protesescu2015}, which has generated much excitement due to their exceptional optical properties.

Shortly following the initial synthesis of perovskite nanocubes, synthesis of perovskite nanoplatelets \cite{Bekenstein2015,Tong2016} was also achieved to further broaden the gamut of applications for perovskite nanocrystals. Compared to their nanocube counterparts, the nanoplatelet geometry offers directional light absorption/emission \cite{Jurow2019} as well as reduced dielectric screening (leading to greatly enhanced exciton binding energies \cite{Wang2018} and radiative recombination rates \cite{Hintermayr2016,Weidman2017}). Recently, these attractive properties have led to intense efforts in applying perovskite nanoplatelets towards a variety of applications such as light-emitting diodes \cite{Peng2019} and photovoltaics \cite{Wei2019}. Understanding how electronic dynamics underlying the photo-physics of perovskite nanocrystals change with nanocrystal geometry is crucial for such practical applications. In particular, perovskite nanoplatelets have been seldom studied at cryogenic temperature to elucidate electron-phonon coupling in the material.

Here, we study CsPbI$_3$ perovskite nanocube and nanoplatelet ensembles at cryogenic temperatures. Absorption spectra reveal an anomalous band-gap shift to higher energies with increasing temperature, which we attribute to band-gap renormalization via electron-phonon coupling. A low-energy absorption tail is also observed in CsPbI$_3$ nanocubes that is likely due to shallow trap states, which implies that iodide perovskite nanocrystals may be less defect-tolerant than their bromide and chloride counterparts at low temperatures.

\section{Experiment}

\begin{figure*}
    \centering
    \includegraphics[width=1\textwidth]{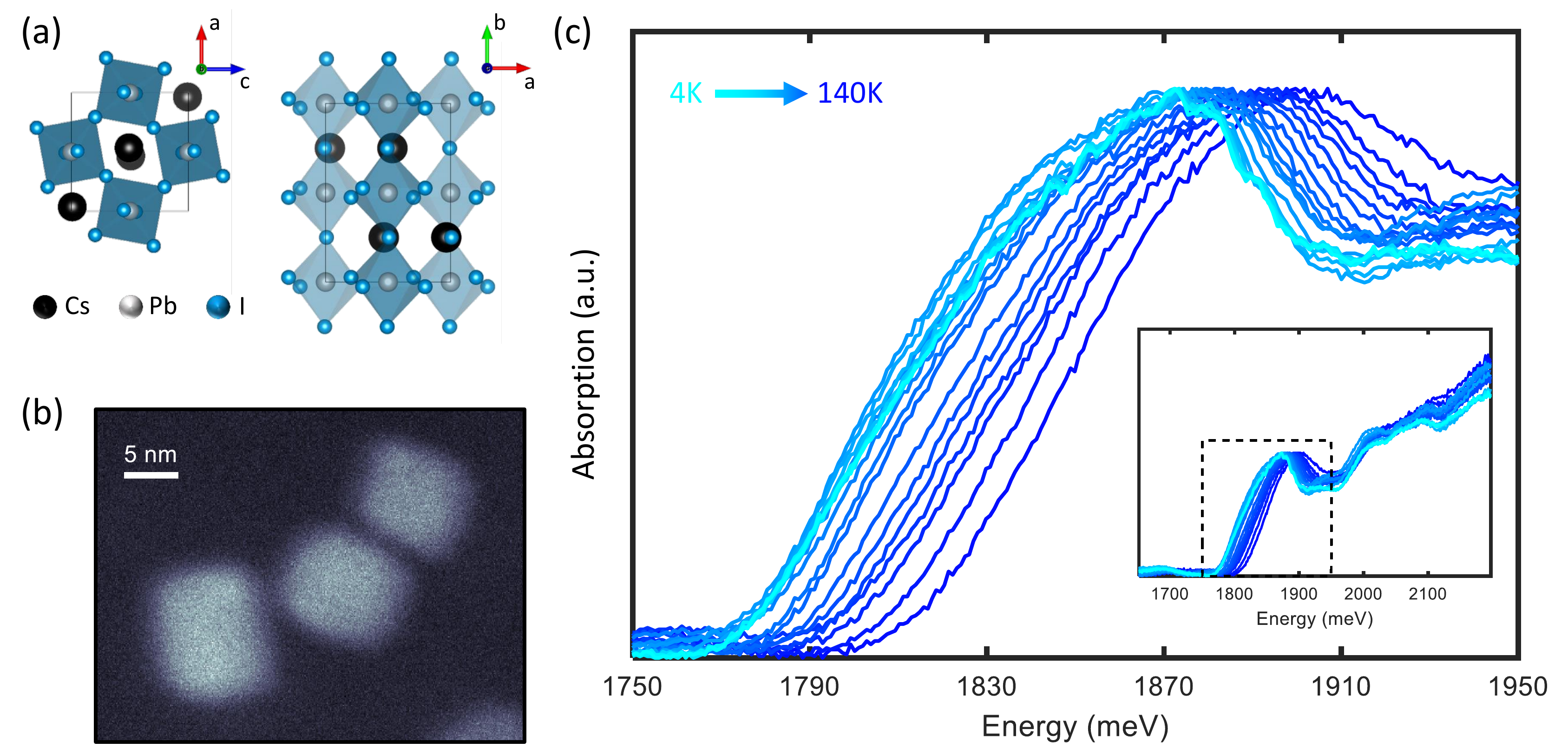}
    \caption{(a) Two perspective views of the orthorhombic perovskite lattice structure of CsPbI$_3$ with axes as shown (plotted using the VESTA software \cite{Momma2011}). The unit cell is denoted by the solid black lines. (b) Transmission electron micrograph of nanocubes. (c) Absorption spectra of CsPbI$_3$ nanocrystals at temperatures ranging from 4 K to 140 K as indicated. The full-range spectra are plotted inset, while the 1S exciton peak outlined by the dashed box is shown in the main plot. The specific temperatures plotted are indicated by the data in Fig.~\ref{Fig2}(b).}
    \label{Fig1}
\end{figure*}

The orthorhombic perovskite lattice structure of the CsPbI$_3$ nanocrystals \cite{Cottingham2016,Bertolotti2017,Sutton2018} is shown in Fig.~\ref{Fig1}(a), and transmission electron micrographs of the nanocubes are shown in Fig.~\ref{Fig1}(b). Measurement of 100 nanocubes informs an average side length of 8.7 $\pm$ 2.6 nm. Although significant size and shape dispersion of the nanoplatelets preclude well-defined average side lengths, their lateral dimensions on the order of tens of nanometers (see Supplemental Information). Their band-gap energy then indicates the out-of-plane thickness to be primarily four polyhedral layers and above.

The nanocubes are synthesized according to the procedures detailed by Protesescu, et al. \cite{Protesescu2015,Protesescu2017}, and the nanoplatelets are synthesized via a method \cite{BonatoPaper} modified from that reported by Sheng, et al. \cite{Sheng2018}. Brief descriptions of each method are detailed in the Supplemental Information.

To study their optical properties at cryogenic temperatures we redisperse the nanocrystals in heptamethylnonane, a branched alkane that forms a transparent glass at cryogenic temperatures \cite{Liu2019}. The colloidal suspension is held in a custom sample holder approximately 0.5 mm thick and mounted in a cold-finger cryostat. Absorption spectra are measured with a broadband white light source and a UV-vis diode array spectrometer.

\section{Results and Discusssion}

CsPbI$_3$ nanocube absorption spectra normalized to the lowest-energy 1S exciton absorption peak at temperatures ranging from 4 K to 140 K are plotted in Fig.~\ref{Fig1}(c). Although multiple peaks are observed that correspond to distinct exciton transitions, here we focus on the 1S exciton absorption peak that reflects the fundamental electronic band-gap (energy-gap) of the nanocrystals. As temperature increases the band-gap exhibits a pronounced blue-shift to higher energies, which is contrary to the red-shift observed in most solids. In the literature, this phenomenon has been referred to as an anomalous band-gap shift \cite{Gobel1998,Choi2001,Yu2011,Saran2017}.

\begin{figure}
    \centering
    \includegraphics[width=0.5\textwidth]{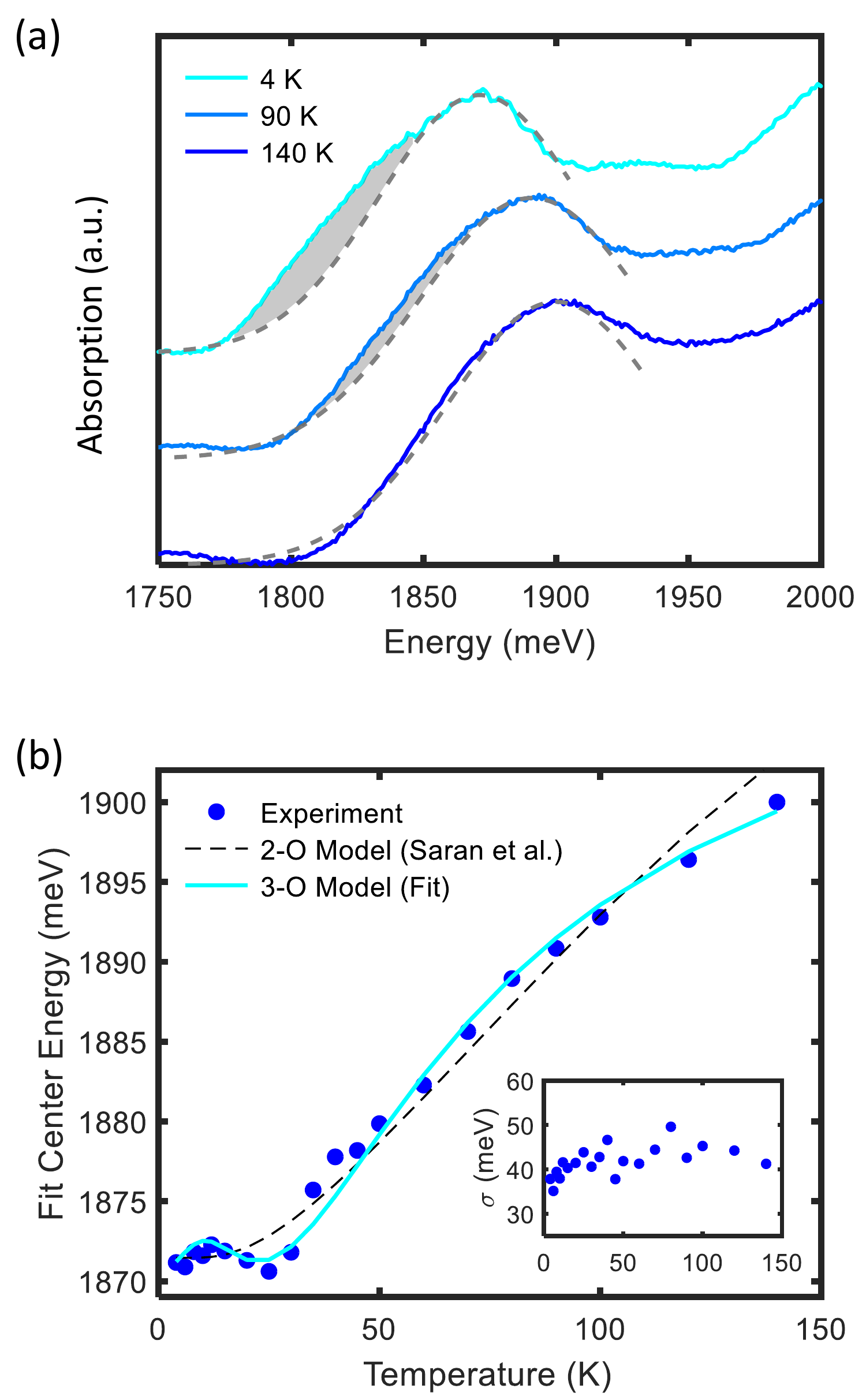}
    \caption{(a) Gaussian peak fits of CsPbI$_3$ absorption spectra at three representative temperatures 4, 90, and 140 K. A low-energy absorption tail, indicated by the shaded gray region, forms at low temperature. (b) Dark-blue dots show fitted Gaussian center energy as a function of temperature, which reflects the material band-gap. A two-oscillator (2-O) model using the fitted parameters from Saran et al. \cite{Saran2017} and a fit to the three-oscillator (3-O) model described in the text are then plotted as the dashed black curve and solid light-blue curve respectively. The fitted Gaussian widths $\sigma$ are plotted inset.}
    \label{Fig2}
\end{figure}

To quantify the band-gap shift, we fit the peaks with Gaussian lineshapes that reflect the size distribution of the nanocrystals. As shown in Fig.~\ref{Fig2}(a), we fit only the top of each peak due to absorption tails present at lower temperatures. The widths $\sigma$ of each Gaussian fit, allowed to vary freely, do not change significantly with temperature (mean width 41.81 meV and standard deviation 3.37 meV). The fitted Gaussian center energies (which agree closely with center energies found from a fourth-order polynomial fit) are plotted in Fig.~\ref{Fig2}(b), which reveals interesting behavior at temperatures below 50 K. Specifically, two clear inflection points at 20 and 30 K are observed that reveal more complicated band-gap behavior than previously reported for photoluminescence measurements of similar perovskite nanocubes \cite{Saran2017}.

The dependence of the electronic band-gap on temperature $T$ may be expressed as \cite{Cardona2005,Yu2011}:
\begin{align}
    E_g(T) = E_0 + AT + \sum\limits_nB_n\left(\frac{1}{e^{\hbar\omega_n/k_BT} - 1} + \frac{1}{2}\right).
\end{align}
The first term $E_0$ is the intrinsic material band-gap at $T = 0$, and the coefficient $A$ in the second term characterizes the change in band-gap due to lattice unit cell expansion/contraction (in the so-called quasi-harmonic approximation \cite{Cardona2005}). Here the change in quantum confinement energy due to expansion/contraction of nanocrystal volume, which we expect to be negligible at low temperatures \cite{Badlyan2019}, is ignored. The third term then represents renormalization of the band-gap due to electron-phonon interactions, where $n$ is summed over all phonon branches and all wave-vectors within the Brillouin zone for each branch. $B_n$ and $\hbar\omega_n$ are the electron-phonon coupling strength and vibrational energy respectively for mode $n$. Whether $B_n$ is positive or negative, resulting in an increase or decrease of the band-gap respectively, arises from a complex interplay of microscopic dynamics and cannot be predicted easily from the properties of a given phonon branch \cite{Garro1996,Gobel1998}. However, accounting for all possible phonon branches throughout the Brillouin zone is often unnecessary in modeling the behavior of real systems. Instead, one \cite{Choi1999} or two \cite{Gobel1998} vibrational modes are usually assumed dominant (referred to as one-oscillator and two-oscillator models) which reduces the summation to either one or two terms respectively.

Here, we find both the one-oscillator and two-oscillator models to be insufficient in modeling the band-gap temperature dependence observed for CsPbI$_3$ nanocubes. As mentioned above, two inflection points are observed that necessitate at least three dominant vibrational modes that independently renormalize the band-gap. A least-squares fit of the band-gap temperature dependence to this three-oscillator model is plotted in Fig.~\ref{Fig2}(b), where good agreement is observed at both high and low temperatures. The fitted parameters are $E_0 = 1916.9$ meV, $A = 0.3$ meV/K, $\hbar\omega_1 = 5.38$ meV, $\hbar\omega_2 = 5.91$ meV, $\hbar\omega_3 = 17.02$ meV, $B_1 = -698.01$ meV, $B_2 = 821.67$ meV, and $B_3 = -217.39$ meV. Instead of the acoustic and optical phonon categories that are usually invoked for two-oscillator models \cite{Yu2011,Saran2017}, a three-oscillator model in perovskite materials align more naturally to the bending, stretching, and rocking perovskite vibrational modes that possess distinct ranges of vibrational energies \cite{Perez-Osorio2015}.

We note that although the two-oscillator model was recently invoked by Saran et al. to model the temperature dependence of photoluminescence center energy in perovskite nanocrystals \cite{Saran2017}, the data points taken at low temperatures (below 50 K) were too sparse to resolve the two inflection points we observe. Their resultant fitted band-gap dependence is plotted in Fig.~\ref{Fig2}b for comparison. In contrast to features in absorption spectra, which are simply proportional to the oscillator strength of each optical transition, features in photoluminescence spectra depend on many other temperature-dependent factors such as the equilibrium fine-structure carrier distribution \cite{Yin2017,Raino2016} and emission Stokes shifts \cite{Qiao2010}. It is therefore unclear whether the apparent two-oscillator behavior of their measurements on CsPbI$_3$ nanocubes was due to coarse-graining effects or the above confounding factors in temperature-dependent photoluminescence.

Lower-energy absorption tails are observed. For ideal nanocubes, the exciton density of states are comprised of delta functions that result in roughly Gaussian absorption peaks (reflecting the nanocrystal size distribution). Absorption tails at lower-energy are therefore indicative of corresponding tails of the electronic density of states, often attributed to impurities \cite{Studenyak2014} or surface states \cite{Guyot-Sionnest2012}. As shown in Fig.~\ref{Fig2}(a), the absorption peak is Gaussian at 140 K and develops a lower-energy tail with decreasing temperature. We attribute this tail to shallow defect states surrounding the valence band-edge that have been shown to arise from lattice point defects \cite{Kang2017}. At high temperatures valence band electrons populate the band-edge in a thermal equilibrium distribution. At low temperatures those electrons then fill the defect states from lowest energy upwards, which comprise a Halperin-Lax type distribution \cite{Halperin1966} with a exp($\sqrt{E}$) dependence \cite{Yan2016,Jean2017}. The disappearance of the tail at 140 K thus suggests a few-meV (comparable to the 140 K Boltzmann energy of 12 meV) defect state energy distribution. Although in principle such defect state absorption should manifest in photoluminescence spectra as well, no clear band-tailing was observed in low-temperature photoluminescence measurements\cite{Saran2017}. This is unsurprising, since above-gap excitation results in competing band-edge and defect state relaxation pathways and emission Stokes shifts (on the order of tens of meV in perovskite nanocrystals \cite{Protesescu2015,Saran2017}) likely differ for defect transitions. For additional comparison, absorption measurements were also performed on CsPbBr$_3$ nanocubes (see Supplementary Material for absorption spectra and synthesis methods). Although a large anomalous band-gap shift was observed (approximately 40 meV from 6 to 140 K), no absorption tail forms at low temperatures.

\begin{figure}
    \centering
    \includegraphics[width=0.5\textwidth]{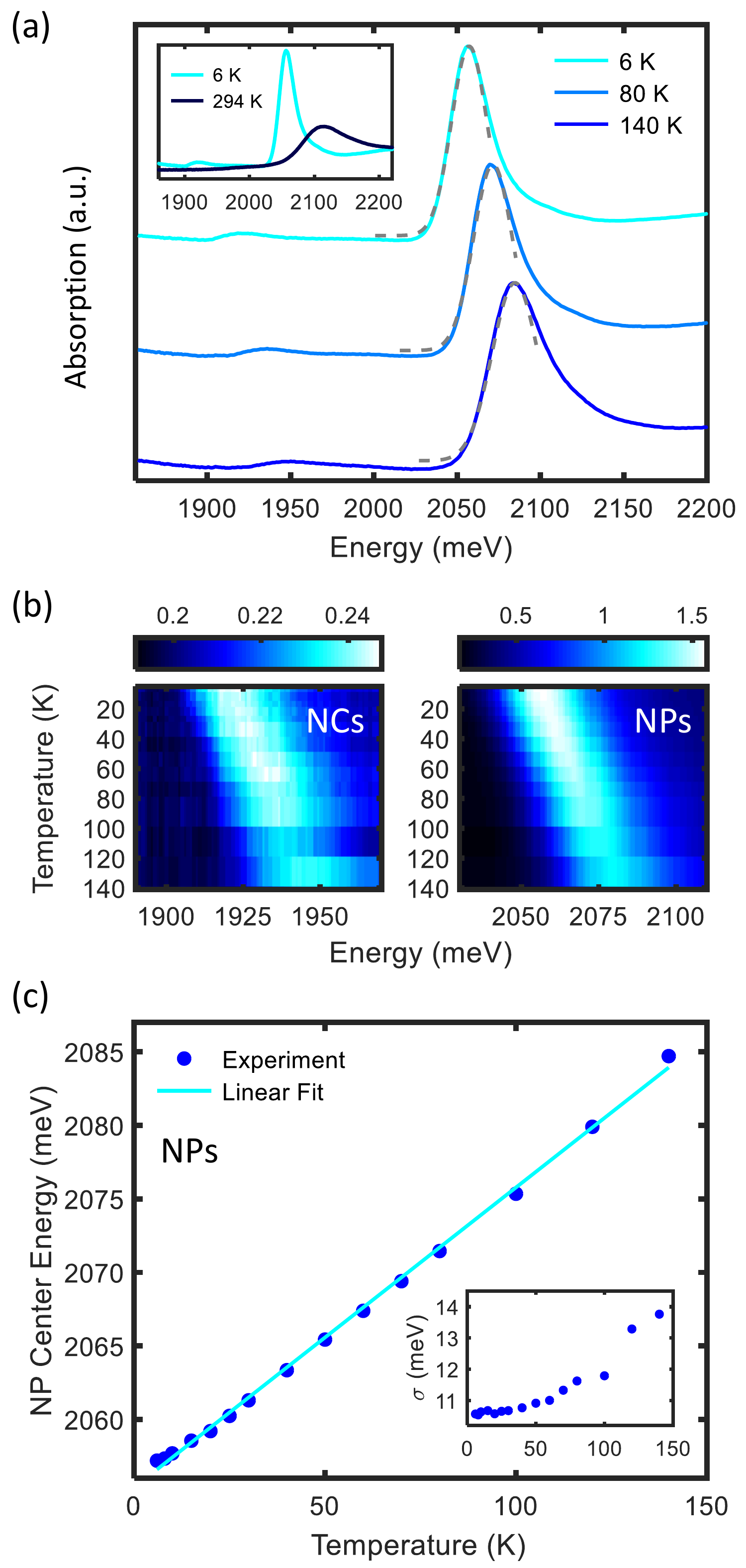}
    \caption{(a) Absorption spectra of CsPbI$_3$ nanoplatelets at three representative temperatures 6, 80, and 140 K. In addition to the main nanoplatelet (NP) absorption peak, a weak nanocube (NC) absorption peak at lower energy appears at low temperatures. Inset shows comparison between 6 K and room-temperature absorption spectra. (b) Optical density surface plots of the NC and NP absorption peaks in (a) as a function of temperature. (c) NP center energies obtained from the absorption peak first moment as a function of temperature, which reflects the material band-gap. A linear fit is plotted as the solid blue curve. The fitted Gaussian widths are plotted inset, which monotonically decrease with decreasing temperature.}
    \label{Fig3}
\end{figure}

Electron-phonon coupling that renormalizes the CsPbI$_3$ bandgap should depend strongly on dimensionality in nanocrystals. In particular, lowering dimensionality should reduce electron-phonon coupling by restricting certain vibrational modes. To investigate the effect of lattice dimensionality on electron-phonon coupling, we repeat the same temperature-dependent absorption measurements on CsPbI$_3$ nanoplatelets. At room-temperature, a single nanoplatelet absorption peak is observed that is blue-shifted relative to the nanocube band-gap due to strong quantum confinement in the out-of-plane direction. At cryogenic temperatures, shown in Fig.~\ref{Fig3}, the absorption spectrum changes in two surprising ways. First, the nanoplatelet absorption peak continues narrowing below 140 K (with no absorption tail), in contrast to the nanocube absorption peak width that remains constant at low temperatures. Second, an additional lower-energy peak also appears with decreasing temperature (see Fig.~\ref{Fig3}a) which, due to its center energy, we attribute to co-synthesized CsPbI$_3$ nanocubes. Temperature-dependence surface plots of both the nanocube and nanoplatelet peaks (measured from the same absorption spectra) are shown in Fig.~\ref{Fig3}b to inform the relative changes in peak optical density.

Again fitting the nanoplatelet absorption peaks to Gaussian lineshapes, the fitted center energies are plotted in Fig.~\ref{Fig3}c. The nearly-linear anomalous band-gap shift indicates weakened electron-phonon interactions and greater importance of band-gap renormalization due to unit cell expansion/contraction with temperature. To quantify these changes, we perform a linear fit of the center energy temperature-dependence. The fitted parameters are $E_0 = 2055.4$ meV and $A = 0.2$ meV/K, where $A$ is comparable to its corresponding nanocube value. Therefore, decreasing dimensionality greatly reduces vibrational band-gap renormalization without strongly affecting that due to changes in unit cell size.

The fitted Gaussian widths $\sigma$, plotted inset in Fig.~\ref{Fig3}c, reveal another interesting aspect of the electronic properties of perovskite nanoplatelets. While the nanocube absorption peak width is approximately constant at cryogenic temperatures, reflecting its inhomogeneously broadened nature, the much narrower nanoplatelet absorption peak exhibits a monotonic decrease in $\sigma$ with decreasing temperature. This indicates that homogeneous broadening in perovskite nanoplatelets contributes even down to cryogenic temperatures. However, a plateau in the linewidth decrease below 50 K, despite homogeneous out-of-plane confinement, reveals an intrinsic ensemble absorption linewidth between 10 and 11 meV. At such energy scales, inhomogeneous broadening due to variation in in-plane confinement of exciton center-of-mass motion, usually considered to be negligible \cite{Nasilowski2016}, could become important. More advanced spectroscopic techniques such as multi-dimensional coherent spectroscopy \cite{Cundiff2013} are needed to disentangle inhomogeneous and homogeneous broadening mechanisms in perovskite nanoplatelets \cite{Liu2019,Liu2019-2}.

\section{Conclusion}

In summary, the absorption of CsPbI$_3$ perovskite nanocrystals are measured at cryogenic temperatures. In addition to the anomalous band-gap shifts to higher energies with increasing temperature, additional inflection points are observed at low temperatures that we attribute to band-gap renormalization by, contrary to a recent study \cite{Saran2017}, three vibrational modes in CsPbI$_3$ nanocubes. Measurement of CsPbI$_3$ nanoplatelets then reveals greatly reduced vibrational band-gap renormalization, which suggests that lowered nanocrystal dimensionality leads to weakened influence of lattice vibrations on electronic dynamics. Lastly, absorption tails are found to form in CsPbI$_3$ nanocubes at low temperatures, which we attribute to defect states surrounding the valence band-edge. While perovskite nanocrystals have been found to be exceptionally defect-tolerant \cite{Huang2017}, our finding suggests that shallow defects may begin to influence the optical properties of iodide nanocubes at cryogenic temperatures. This work motivates further study of electron-phonon coupling in perovskite nanocrystals to minimize their deleterious effects.

\begin{acknowledgments}
This work was supported by the Department of Energy grant number DE-SC0015782 and by the Sao Paulo Research Foundation, under the grant number 2013/16911-2. D.B.A. and G.N. acknowledge support by fellowships from the Brazilian National Council for Scientific and Technological Development (CNPq). Research was also supported by LNNano/CNPEM/MCTIC, where the TEM measurements were performed.
\end{acknowledgments}

\end{document}